\documentclass[a4paper,12pt,oneside]{article}
\usepackage[cp1250]{inputenc}

\usepackage{ a4, fancyhdr, indentfirst, graphicx, amsmath, amsfonts}

\newtheorem{theorem}{Theorem}

\newtheorem{proposition}{Proposition}

\newtheorem{corollary}{Corollary}

\begin{document}

\title{On the key exchange with nonlinear polynomial maps of stable
degree}
\author{V. Ustimenko, A. Wroblewska\\
University of Maria Curie Sklodowska, Lublin, Poland\\
\texttt{ustymenko\_vasyl@yahoo.com,
awroblewska@hektor.umcs.lublin.pl }}
\date{\today}
\maketitle

\begin{abstract}
We say that the sequence $g_n$, $n\ge 3$, $n \rightarrow \infty$ of
polynomial transformation bijective maps of free module $K^n$ over
commutative ring $K$ is a sequence of stable degree if the order of
$g_n$ is growing with $n$ and the degree of each nonidentical
polynomial map of kind ${g_n}^k$ is an independent constant $c$. A
transformation $b={\tau}
  {g_n}^k {\tau}^{-1}$, where $\tau$ is affine bijection,  $n$ is large
and $k$ is relatively small, can be used as a base of group
theoretical Diffie-Hellman key exchange algorithm for the Cremona
group $C(K^n)$ of all regular automorphisms of $K^n$. The specific
feature of this method is that the order of the base may be unknown
for the adversary because of the complexity of its computation. The
exchange can be implemented by tools of Computer Algebra (symbolic
computations). The adversary can not use the degree of righthandside
in $b^x=d$ to evaluate unknown $x$ in this form for the discrete
logarithm problem.

In the paper we introduce the explicit constructions of sequences of
elements of stable degree for cases $c=3$ for each commutative ring
$K$ containing at least 3 regular elements and discuss the
implementation of related key exchange and public key algorithms.


\end{abstract}

\section{Introduction}
Discrete logarithm problem can be formulated for general finite
group $G$. Find a positive integer $x$ satisfying condition $g^x=b$
where $g \in G$ and $b \in G$. The problem has reputation to be a
difficult one. But even in the case of cyclic group $C$ there are
many open questions. If $C=Z_{p-1}^*$ or $C=Z_{pq}^*$ where $p$ and
$q$ are "sufficiently large" primes then the complexity of discrete
logarithm problem justify classical Diffie-Hellman key exchange
algorithm and RSA public key encryption, respectively. In most of
other cases complexity of discrete logarithm problem is not
investigated properly. The problem is very dependent on the choice
of the base $g$ and the way of presentation the data on the group.
Group can be defined via generators and relations, as automorphism
group of algebraic variety, as matrix group, as permutation group
etc. In this paper we assume that $G$ is a subgroup of $S_{p^n}$
which is a group of polynomial bijective transformation of vector
space ${F_p}^n$ into itself. Obviously $|S_{p^n}|=(p^n)!$, it is
known that each permutation $\pi$ can be written in the form $x_1
\rightarrow f_1(x_1, x_2, \dots x_n), x_2 \rightarrow f_2(x_1, x_2,
\dots x_n),\dots,x_n \rightarrow f_n(x_1, x_2, \dots x_n)$, where
$f_i$ are multivariable polynomials from $F_p[x_1, x_2, \dots,
x_n]$. The presentation of $G$ as a subgroup of $S_{p^n}$ is chosen
because the Diffie-Hellman algorithm here will be implemented by the
tools of symbolic computations. Other reason is universality, as it
follows from classical Cayley results each finite group $G$ can be
embedded in $S_{p^n}$ for appropriate $p$ and $n$ in various ways.

Let $F_p$, where $p$ is prime, be a finite field. Affine
transformations ${\rm x} \rightarrow A{\rm x}+b$, where $A$ is
invertible matrix and $b \in (F_p)^n$, form an affine group
$AGL_n(F_p)$ acting on ${F_p}^n$.

   Affine transformations
form an affine group $AGL_n(F_p)$ of order $p^n(p^n-1)(p^n-p) \dots
(p^n-p^{n-1})$ in the symmetric group $S_{p^n}$ of order $p^n!$. In
\cite{Moore} the maximality of $AGL_n(F_p)$ in $S_{p^n}$ was proven.
So we can present each permutation $\pi$ as a composition of several
"seed" maps of kind $\tau_1 g \tau_2$, where $\tau_1, \tau_2 \in
AGL_n(F_p)$ and $g$ is a fixed map of degree $\ge 2$.

We can choose the base of ${F_p}^n$ and write each permutation $g\in
S_{p^n}$ as a "public rule":

$x_1 \rightarrow g_1(x_1, x_2, \dots, x_n), x_2 \rightarrow g_2(x_1,
x_2, \dots, x_n), \dots, x_n \rightarrow g_n(x_1, x_2, \dots, x_n)$.

Let $g^k \in S_{p^n}$ be the new public rule obtained via iteration
of $g$. We consider Diffie-Hellman algorithm for $S_{p^n}$ for the
key exchange in the case of group . Correspondents Alice and Bob
establish $g\in S_{p^n}$ via open communication channel, they choose
positive integers $n_A$ and $n_B$, respectively. They exchange
public rules $h_A=g^{n_A}$ and $h_B=g^{n_B}$ via open channel.
Finally, Alice and Bob compute common transformation $T$ as
${h_B}^{n_A}$ and ${h_A}^{n_B}$, respectively.

In practice they can establish common vector $ v=(v_1, v_2, \dots,
v_n)$, $v_i\in F_p,~~i=1,\dots,n$ via open channel and use the
collision vector $T(v)$ as a password for their private key
encryption algorithm.

This scheme of symbolic Diffie-Hellman algorithm can be secure, if
the order of $g$ is "sufficiently large" and adversary is not able
to compute number $n_A$ (or $n_B$) as functions from degrees for $g$
and $h_A$. Obvious bad example is the following: $g$ sends $x_i$
into ${x_i}^t$ for each $i$. In this case  $n_A$ is just a ratio of
${\rm deg}h_A$ and ${\rm deg} g$.

To avoid such trouble one can look at family of subgroups $G_n$ of
$S_{p^n}$, $n\rightarrow \infty$ such that maximal degree of its
elements equal $c$, where $c$ is small independent constant (groups
of degree $c$ or groups of stable degree). Our paper is devoted to
explicit constructions of such families.

We refer to a sequence of elements $g_n\in G_n$ such that all its
nonidentical powers are of degree $c$ as element of stable degree.
This is equivalent to stability of families of cyclic groups
generated by $g_n$. Of course, cyclic groups are important for the
Diffie-Hellman type protocols.

It is clear that affine groups $AGL_n(F_p)$, $n \rightarrow \infty$
form a family of subgroups of stable degree for $c=1$ and all
nonidentical affine transformations are of stable degree. Notice
that if $g$ is a linear diagonalisable element of $AGL_n(F_p)$, then
discrete logarithm problem for base $g$ is equivalent to the
classical number theoretical problem. Obviously, in this case we are
losing the flavor of symbolic computations.
One can take a subgroup $H$ of $AGL_n(F_p)$ and consider its
conjugation with nonlinear bijective polynomial map $f$. Of course
the group $H'=f^{-1}Hf$ will be also a stable group, but for "most
pairs" $f$ and $H$  group $H'$ will be of degree ${\rm deg f} \times
{\rm deg} f^{-1} \ge 4$ because of nonlinearity $f$ and $f^{-1}$.

So the problem  of construction an infinite families of subgroups
$G_n$ in $S_{p^n}$ of degree 2 and 3 may attract some attention.

  General problem of construction an infinite families of
stable subgroups $G_n$ of $S_{p^n}$ of degree $c$ satisfying some
additional conditions (unbounded growth of minimal order of
nonidentical  group elements, existence of well defined projective
limit, etc) can be also interesting because of possible applications
in cryptography.

Notice that even we conjugate nonlinear $C$ with invertible linear
transformation $\tau \in AGL_n(F_p)$, some of important
cryptographical parameters of $C$ and $C'= {\tau}^{-1}C\tau$ can be
different. Of course conjugate generators $g$ and $g'$ have the same
number of fixed points, same cyclic structure as permutations, but
counting of equal  coordinates for pairs  ($x$, $g(x)$) and ($x$ ,
$g'(x)$) may bring very different results.

So two conjugate families of stable degree are not quite equivalent
because corresponding cryptoanalitical problems may have different
complexity.

We generalize the above problem for the case of Cremona group of the
free module $K^n$, where $K$ is arbitrary commutative ring $K$. For
the  cryptography case of finite rings is the most important.
  Finite field $F_{p^n}$, $n \ge 1$ and cyclic rings $Z_m$ (especially
$m=2^7$ ( ASCII codes), $m=2^8$ (binary codes), $m=2^{16}$
(arithmetic), $m=2^{32}$ ( double precision arithmetic)) are
especially popular. Case of infinite rings $K$ of characteristic
zero (especially $Z$ or $C$) is an interesting as well because of
Matijasevich multivariable prime approximation polynomials can be
defined there (see, for instance \cite{Nato} and further
references).

  So it is natural to change
a vector space ${F_p}^{n}$ for free module $K^n$ (Cartesian power of
$K$) and the family and symmetric group $S_{p^n}$ for Cremona group
$C(n,K)$ of all polynomial automorphisms of $K^n$.

We repeat our definition for more general situation of commutative
ring.

Let $G_n$, $n\ge 3$, $ n \rightarrow \infty$ be a sequence of
subgroups of $C(n,K)$. We say that $G_n$ is a family of groups of
stable degree (or subgroup of degree $c$) if the maximal degree of
representative $g\in G_n$ is some independent constant $c$.

Recall, that cases of degree 2 and 3 are especially important.

The first family of stable subgroups of $C_n(F_q)$, $K=F_q$ with
degree 3 was practically established in \cite{Wrob}, where the
degrees of polynomial graph based public key maps were evaluated.
But group theoretical language was not used there and  the problem
of the key exchange was not considered.

So we reformulate the results of \cite{Wrob} in terms of Cremona
group over a general ring in section 2 of current paper.

Additionally we show the existence of cubic elements of large order
in case of finite field.

Those results are based on the construction of the family $D(n, q)$
of graphs with large girth and the description of their connected
components $CD(n, q)$. The existence of infinite families of graphs
of large girth had been proven by Paul Erd{\"o}s' (see
\cite{Bollobas}). Together with famous Ramanujan graphs introduced
by G. Margulis \cite{Margulis} and investigated in \cite{Raman}
graphs $CD(n,q)$ is one of the first explicit constructions of such
a families with unbounded degree. Graphs $D(n, q)$ had been used for
the construction of LDPS codes and turbocodes which were used in
real satellite communications (see \cite{Lodge1}, \cite{Lodge2},
\cite{Kim}, \cite{Klisow}), for the development of private key
encryption algorithms \cite{Coord},\cite{Cryptim},
\cite{Arcs},\cite{Kotor}, the option to use them for public key
cryptography was considered in \cite{Max}, \cite{Varna} and in
\cite{Ling} , where the related dynamical system had been introduced
(see also surveys \cite{Directed}, \cite{Nato}).

The computer simulation show that stable subgroups related to $D(n,
q)$ contain elements of very large order but our theoretical linear
bounds on the order are relatively weak. We hope to improve this gap
in the future and justify the use of $D(n, q)$ for the key exchange.

In section 4 we also will use graphs and related finite automata for
the constructions of families of stable subgroups with degree 3 of
Cremona group $C(n,K)$ over general ring $K$ containing elements of
large order (order is growing with the growth of $n$).
  First family of stable
groups were obtained via studies of simple algebraic graphs defined
over $F_q$. For general constructions of stable groups over
commutative ring $K$ we use directed graphs with the special
colouring. The main result of the paper is the following statement.

\begin{theorem}
For each commutative ring $K$ with at least 3 regular elements there
is a families  $Q_n$ of
  Cremona group $C(K^n)$ of degrees
3 such that the projective limit $Q$ of $Q_n$, $n \rightarrow
\infty$ is well defined, the group $Q$ is of infinite order, it
contains elements $g$ of infinite order,
  such
that there exists
  a sequence $g_n\in Q_n$
  $n \rightarrow \infty$
of stable elements such that ${\rm lim}g_n=g$.
\end{theorem}


The family $Q_n$ is obtained via explicit constructions. So we may
use in the finite ring $K$ with at least 3 regular
  elements
the sequence equivalent to $g_n$ for the key exchange. We show that
the growth of the order of $g_n$ when $n$ is growing can be bounded
from below by some linear function ${\alpha}\times n + \beta$. In
case of such a sequence of groups $G_n=Q_n$  we can modify a
sequence $g_i$ of elements of stable degree  by conjugation with
$h_i \in G_i$. New sequence $d_i={h_i}^{-1}g_i{h_i}$ can be also a
sequence of elements of stable degree.

Let us discuss the asymmetry of our modified Diffie-Hellman
algorithms of the key exchange in details. Correspondents Alice and
Bob are in different shoes. Alice chooses dimension $n$, element
$g_n$ as in theorem above, element $h \in Q_n$ s
  and
affine transformation $\tau \in AGL_n(K)$. So she obtains the base
$b={\tau}^{-1}h^{-1}g_nh\tau$ and sends it in the form of standard
polynomial map to Bob.

Our groups $Q_n$ are defined by the set of their generators and
Alice can compute words $h^{-1}g_nh$, $b$ and its powers very fast.
So Alice chooses rather large number $n_A$ computes $c_A=b^{n_A}$
and sends it to Bob. At his turn Bob chooses own key $n_B$ computes
$c_B=b^{n_B}$. He and Alice are getting the collision map $c$ as
${c_A}^{n_B}$ and  ${c_B}^{n_A}$ respectively.

{\em Remark.} Notice that the adversary is in the same shoes with
public user Bob. He (or she) need to solve one of the equations
$b^x=c_B$ or $b^x=c_A$. The algorithm is implemented in the cases of
finite fields and rings $Z_m$ for family of groups  $Q_n$. We
present its time evaluation (generation of $b$ and $b^n_A$ by Alice
and computation of $b^c_B$ by Bob) in the last section of paper. We
continue studies of orders of $g_i$ theoretically and by computer
simulation.

The computer simulation show that the number of monomial expressions
of kind $x^{i_1}x^{i_2}x^{i_3}$ with nonzero coefficient is rather
close to binomial coefficient ${C_n}^3$. So the time of computation
$b^{n_B}$, ${c_B}^{n_A}$ and ${c_A}^{n_B}$ can be evaluated via the
complexity of computation of the composition of several general
cubical polynomial maps in $n$ variable.

\section{Walks on infinite forest $D(q)$ and corresponding groups}

\subsection{Graphs and incidence system}

The missing definitions of graph-theoretical concepts which appear
in this paper can be found in \cite{Bollobas}. All graphs we
consider are simple, i.e. undirected without loops and multiple
edges. Let $V(G)$ and $E(G)$ denote the set of vertices and the set
of edges of $G$, respectively. Then $|V(G)|$ is called the {\em
order} of $G$, and $|E(G)|$ is called the {\em size} of $G$. A path
in  $G$ is called {\em simple} if all its vertices are distinct.
When it is convenient, we shall identify $G$ with the corresponding
anti-reflexive binary relation on $V(G)$, i.e. $E(G)$ is a subset of
$V(G) \times V(G)$ and write $vGu$ for the adjacent vertices $u$ and
$v$ (or neighbors). The sequence of distinct vertices $v_1, \dots ,
v_t$, such that  $v_i G v_{i+1}$ for $i=1, \dots , t-1$ is the pass
in the graph. The length of a  pass is a number of its edges. The
distance ${\rm dist}(u, v)$ between two vertices is the length of
the shortest pass  between them. The diameter of the graph is the
maximal distance between two vertices $u$ and $v$  of the graph. Let
$C_m$ denote the cycle of length $m$ i.e. the sequence of distinct
vertices $v_1, \dots, v_m$ such that $v_i Gv_{i+1}$, $i=1, \dots ,
m-1$ and $v_m Gv_1$.
   The girth of a graph $G$,
denoted by $g=g(G)$, is the length of the shortest cycle in $G$. The
degree of vertex $v$ is the  number of its neighbors (see
\cite{Biggs} or \cite{Bollobas}).

  The
incidence structure is the set $V$ with partition sets $P$ (points)
and $L$ (lines) and symmetric binary relation $I$ such that the
incidence of two elements implies that one of them is a point and
another is a line. We shall identify $I$ with the simple graph of
this incidence relation (bipartite graph). If number of neighbours
of each element is finite and depends only
  on its type (point or line), then the incidence structure is a tactical
configuration in the sense of Moore (see \cite{Moore}).
The graph is $k$-regular if each of its vertex has degree $k$, where
$k$ is a constant. In this section we reformulate results of
\cite{Comp}, \cite{Expl} where the $q$-regular tree was described in
terms of equations over finite field $F_q$.

Let $q$ be a prime power, and let $P$ and $L$ be two
  countably infinite dimensional vector spaces
over $F_q$. Elements of $P$ will be called {\it points} and those of
$L$ {\it lines}.
  To distinguish
points from lines we use parentheses and brackets: If $x\in V$, then
$(x)\in P$ and $[x]\in L$. It will also be advantageous to adopt the
notation for coordinates of points and lines introduced in
\cite{Margulis}:
$$(p)=(p_1, p_{11},p_{12},p_{21},p_{22},p'_{22},p_{23},\ldots,
p_{ii},p'_{ii},p_{i,i+1},p_{i+1,i},\ldots),$$
$$[l]=[l_1,l_{11},l_{12},l_{21},l_{22},l'_{22},l_{23},\ldots,
l_{ii},l'_{ii},l_{i,i+1},l_{i+1,i},\ldots).$$
\smallskip

We now define an incidence structure $(P, L, I)$ as follows. We say
the point $(p)$ is incident with the line $[l]$, and we write
$(p)I[l]$, if the following relations between their coordinates
hold:
\smallskip
        $$l_{11}-p_{11}=l_1p_1$$
        $$l_{12}-p_{12}=l_{11}p_1$$
        $$l_{21}-p_{21}=l_1p_{11} \eqno(1)$$
        $$l_{ii}-p_{ii}=l_1p_{i-1,i}$$
        $$ l'_{ii}-p'_{ii}=l_{i,i-1}p_1$$
        $$l_{i,i+1}-p_{i,i+1}=l_{ii}p_1$$
        $$l_{i+1,i}-p_{i+1,i}=l_1p'_{ii}$$
\smallskip
\noindent (The last four relations are defined for $i\ge 2$.) This
incidence structure $(P, L, I)$ we denote as $D(q)$. We speak now of
  the {\it incidence graph} of $(P, L, I)$, which has
the vertex set $P\cup L$ and edge set consisting of all pairs
$\{(p),[l]\}$ for which $(p)I[l]$.

To facilitate notation in future results, it will be convenient for
us to define $p_{-1,0} = l_{0,-1} =p_{1,0} = l_{0,1} = 0$, $p_{0,0}
= l_{0,0} =-1$, $p'_{0,0} = l'_{0,0} = 1$, $p_{0,1} = p_1$, $l_{1,0}
= l_1$,   $l'_{1,1} = l_{1,1}$,  $p'_{1,1} = p_{1,1}$, and to
rewrite (1) in the form :

$$l_{ii}-p_{ii}=l_1p_{i-1,i}$$
$$l'_{ii}-p'_{ii}=l_{i,i-1}p_1$$
$$l_{i,i+1}-p_{i,i+1}=l_{ii}p_1$$
$$l_{i+1,i}-p_{i+1,i}=l_1p'_{ii}$$

         for $i=0,1,2,\ldots $

Notice that for $i = 0$, the four conditions (1) are satisfied by
every point and line, and, for $i=1$, the first two equations
coincide and give $l_{1,1}-p_{1,1}=l_1p_1$.

For each positive integer $k\ge 2$
  we obtain an incidence structure
$(P_k, L_k, I_k)$ as follows.  First, $P_k$ and $L_k$ are obtained
from $P$ and $L$, respectively,
  by simply projecting each vector onto its
  $k$ initial coordinates.
The incidence $I_k$ is then defined by imposing the first $k\!-\!1$
incidence relations and ignoring all others. For fixed $q$, the
incidence graph corresponding to the structure $(P_k, L_k, I_k)$ is
denoted by $D(k, q)$. It is convenient to define $D(1, q)$ to be
equal to $D(2,q)$. The properties of the graphs $D(k,q)$ that we are
concerned with described in the following proposition.
\bigskip
\begin{theorem} \cite{Expl} Let $q$ be a prime power, and
$k\ge 2$. Then

(i) $D(k,q)$ is a $q$-regular edge-transitive bipartite graph of
  order $2q^k$ ;

(ii) for odd $k$, $g(D(k, q)) \geq k+5$, for even $k$, $g(D(k, q))
\geq k+4$

\end{theorem}

\bigskip

    We have a natural one to one correspondence between the
coordinates  2,3, $\ldots$, $n$, $\dots$ of  tuples (points or
lines) and equations. It is convenient for us  to  rename  by $i+2$
the  coordinate  which corresponds to  the  equation  with the
number  $i$  and write
  $[l] = [l_1, l_2, \ldots, l_n, \ldots]$ and
  $(p)=(p_1, p_2, \ldots, p_n, \ldots)$ (line and point in
''natural coordinates'').

      Let $\eta_i$ be the map ''deleting all coordinates with
numbers $>i$''
  from $D(q)$ to $D(i, q)$, and $\eta_{i, j}$ be map "deleting all
coordinates with numbers $>i$ " from $D(j, q)$, $j >i$ into $D(i,
q)$.

        The following statement follows directly from the definitions:

     \begin{proposition} (see, \cite{Expl}) The projective
limit of $D(i, q),  \eta _{i,j}$, $ i \rightarrow \infty$ is an an
infinite forest $D(q)$.

\end{proposition}

Let us consider the description of connected components of the
graphs.

    Let $k \ge 6$, $t=[(k+2)/4]$, and let $u = (u_1, u_{11}, \cdots,
u_{tt}, u^{\prime}_{tt}, u_{t, t+1}, u_{t+1, t}, \cdots)$ be a
vertex of $D(k, q)$. (It does not matter whether $u$ is a point or a
line). For every $r$, $2 \le r \le t$, let

$a_r = a_r(u)=\displaystyle{\sum_{i=0}^{m}  (u_{ii}u'_{r-i, r-i} -
u_{i, i+1}u_{r-i, r-i-1})}$,

and $a=a(u)=(a_2, a_3, \cdots, a_t)$. (Here we define

$p_{-1, 0} = l_{0, -1} = p_{1, 0} = l_ {0, 1} = 0$, $p_{00}=l_{00}=
-1$, $p_{0, 1} = p_1$, $l_{1, 0} = l_1$, $p'_{00}= l'_{00}=1$
$l'_{11}=l_{11}$, $p'_{1, 1}=p_{1, 1}$).

   In \cite{Comp} the following statement was proved.

\begin{proposition}  Let $u$ and $v$ be vertices from the
same component of $D(k, q)$. Then $a(u) = a(v)$. Moreover, for any
$t-1$ field elements $x_i \in F_q$, $2\le  t  \le [(k+2)/4]$, there
exists a vertex $v$ of $D(k, q)$ for which

        $a(v) = (x_2, \ldots, x_t)=(x)$.

\end{proposition}

       Let us consider the following
equivalence relation $\tau$ : $u \tau v$ iff $a(u)=a(v)$ on the set
$P\cup L$ of vertices of $D(k, q)$ ($D(q)$). The equivalence class
  of $\tau$ containing the vertex $v$ satisfying $a(v)=(x)$  can be
  considered as the set
of vertices for the induced subgraph $EQ_{(x)}(k, q)$
($EQ_{(x)}(q)$) of the graph $D(k, q)$ (respectively, $D(q)$). When
$(x)=(0, \cdots, 0)$, we will omit the index $v$ and write simply
$EQ(k, q)$.

Let $CD(q)$ be the connected component of $D(q)$ which contains $(0,
0, \ldots )$.
  Let $\tau'$ be an
equivalence relation on $V(D(k, q))$ ($V(D(q))$) such that the
equivalences classes are the totality of connected components of
this graph. Obviously  $u\tau{v}$ implies $u\tau '{v}$. If {\rm char
$F_q$
  is an odd number, the converse of the last proposition is true
  (see \cite{Nato} and further references).

\begin{proposition} Let $q$ be an odd number.
Vertices $u$ and $v$ of $D(q)$ ($D(k, q))$ belong to the same
connected component iff $a(u) = a(v)$, i.e., $\tau=\tau'$ and
$EQ(q)=CD(q)$ ($EQ(k, q)=CD(k, q)$).
\end{proposition}

    The condition $char F_q \ne 2$ in the last proposition is
  essential.
For instance, the graph $EQ(k, 4))$, $k>3$, contains 2 isomorphic
connected components. Clearly $EQ(k, 2)$ is a union of cycles $CD(k,
2)$.
  Thus neither
$EQ(k, 2)$ nor $CD(k, 2)$ is an interesting family of graphs of high
girth. But the case of graphs $EQ(k, q)$, $q$ is a power of 2, $q >
2$ is very important for coding theory.

\begin{corollary}

      Let us consider a general vertex
    $$x = (x_1, x_{1, 1}, x_{2, 1}, x_{1, 2} \cdots , x_{i, i}, x_{i,
i}^{'},  x_{i+1, i}, x_{i, i+1},
  \cdots),$$ $i=2, 3, \cdots$ of
the connected component $CD(k, F_q)$, which contains a chosen vertex
$v$. Then coordinates  $x_{i, i}$, $x_{i, i+1}$, $x_{i+1,i}$ can be
chosen independently as ``free parameters'' from $F_q$ and
$x^{\prime}_{i,i}$ could be computed successively as the unique
solutions of the equations $a_i(x) = a_i(v)$, $i=1, \ldots$.
\end{corollary}

\subsection{Geometrical interpretation of the algorithm}
We can change $F_q$ for the integral domain $K$ and introduce the
graph $D(K)$ as the graph given by equations (1) over $K$ and repeat
all results of the previous section. If we assume that $K$ is the
general commutative ring then we will lose just the bounds on the
girth.

The graph $D(K)$, where $K$ is integral domain  is a forest
consisting of isomorphic edge-transitive trees (see \cite{Coord} or
\cite{Ling}).

Notice that each tree is a bipartite graph. We may choose a vertex
${\rm x}$ and refer to all vertices on even distance from it as
points. So all remaining vertices are lines.

We may identify all vertices from $P=K^{\infty}$ with the union of
point-sets for all trees from $D(K)$. Another copy $L$ of
$K^{\infty}$ we will treat as totality of all lines in our forest.

For our Diffie-Hellman key exchange protocol Alice has to go to
infinite magic forest $D(K)$ and do the following lumberjack's
business

1) Truncate all trees there by deleting all components with number
$\ge n+1$. So Alice gets a finite dimensional graph $D(n, K)$ which
is a union of isomorphic connected components $CD(n, K)$- truncated
trees.

Notice, if you plant a truncated tree $CD(n, K)$ and let $n
\rightarrow \infty$ then it will grow to a projective limit of
$CD(n,K)$, which is an infinite regular tree.

2) We define a special colouring of graph $D(n, K)$ (or $D(K)$) in
the following way. Let us identify our simple graph with the
directed graph of corresponding symmetric binary relation. We
introduce  the colour of the directed  arrow between two ordered
vertices of our graph  $v_1$ and $v_2$ as the difference of their
first coordinates. It is $l_{0, 1}-p_{0,1}$ if $v_1$ is a point
$(p)$ and $-(l_{0, 1}-p_{0,1})$ if $v_1$ is a line $[l]$.

Let $X(\alpha, \beta)$ be the operator on the vertices of the graph
$D(K)$ moving point $(p)$ to its neighbor alongside the edge of
colour $\alpha$ and moving line $l$ to its neighbor alongside the
edge of colour $\beta$. It is clear that $X(\alpha, \beta) X(-\beta,
-\alpha)$ is an identity map $e$. So $X(\alpha, \beta)^{-1}=
X(-\beta, -\alpha)$. We assume,  that $N_{\alpha}=X(\alpha,
\alpha)$.

Let us define the infinite group $GD(K)$ generated by elements of
kind $g=N_{\alpha_1}N_{\alpha_2}\dots
  N_{\alpha_{2s-1}}N_{\alpha_{2s}}({\rm x})$, $s=1,2 \dots$
corresponding  to walks of even length within the tree starting in
the general vertex ${\rm x}$. It is a transformation group of
variety $P \cup L$. It acts transitively on $P$ (or $L$). $(GD(K),
P)$ is a subgroup of Cremona group for variety $K^{\infty}$.

The computation of $g=N_{\alpha_1}N_{\alpha_2} \dots
N_{\alpha_{2s-1}}N_{\alpha_{2s}}({\rm x})$ in the transformation
group $(GD(K), P)$ corresponds to walk in $D( K)$ of even length
within the tree starting with the point ${\rm x}$. So the group $G$
is the totality of all point to point walks in our forest.

The composition of $g_1$ and $g_2$ from variable ${\rm x}$ is the
walk corresponding to $g_1$ with starting point ${\rm x}$ combined
with the walk corresponding to $g_2$ with the starting point
$g_1({\rm x})$ and final point $g_2(g_1({\rm x}))$,

Each pass of even length in the graph starting from a point $(p)$
can be obtained as a sequence $(p)$,  $v_1=N_{\alpha_1}(p), v_2=
N_{\alpha_2}(v_1), \dots, v_{2k}=N_{\alpha_{2k}}(v_{2k-1})$.

Each element of $GD(K)$ has an infinite order because our forest
does not contain cycles.

Let us consider our symbolic Diffie -Hellman protocol for the
infinite  transformation group ${GD(K), P}$.

a) In case of this group  Alice is hiding a general point ${\rm x}$
by "quasi random" affine transformation $T$ and sending $g(T({\rm
x}))$ to Bob.

b) Further  Bob chooses his key ${k_B}$ and  computes transformation
$h_b=g(T(x))^{k_B}$ of point set for the tree. He makes this
computation root in "darkness" because he has no information on the
forest, he has to apply standard tools for symbolic computations.

c) Alice computes $h_A=g(T(x))^{k_A}$. She can make it fast because
via the repetition of  the walk $g$ from the vertex $T(x)$ several
times.

d) Alice and Bob are getting the collision vector as ${h_B}^k_A$ and
${h_A}^k_B$ respectively.

\subsection{Truncated trees and corresponding stable group}
Now we change the forest $D(K)$ on the bunch of truncated trees from
$D(n,K)$.
  Computation $g=N_{\alpha_1}N_{\alpha_2} \dots
N_{\alpha_{2s-1}}N_{\alpha_{2s}}({\rm x})$  generate the group
$(GD(n,K), P\cup L)$ corresponding all walks in $D(n, K)$ of even
length starting in vertex ${\rm x}$.

Each pass of even length in the graph starting from a point $(p)$
can be obtained as a sequence $(p)$,  $v_1=N_{\alpha_1}(p), v_2=
N_{\alpha_2}(v_1), \dots, v_{2k}=N_{\alpha_{2k}}(v_{2k-1})$.

 Now Alice and Bob can do the key exchange similarly to the case
 of $GD(K)$ but in finite group $GD(n, K)$, where $K$ is a finite ring

{\rm REMARK}. The generalised graph $D(n, K)$ can be defined on the
vertex set $K^n \cup K^n$ in case of arbitrary ring $K$ by equations
(1).  Notice that if $K$ contains zero divisors then girth is
dropping, it is bounded by constant.

 The next result follows instantly from \cite{Wrob} .

\begin{theorem}
Let $K$ be a commutative ring containing at least 3 regular
elements. Sequence of subgroups $GD(n,K)$ of Cremona group $C(n,K)$
form a family of stable subgroups of degree 3.

\end{theorem}

We refer to element $g=N_{\alpha_1}N_{\alpha_2} \dots
N_{\alpha_{2s-1}}N_{\alpha_{2s}}$ for which $\alpha_i \ne
\alpha_{i+1}$, $i=1,2 \dots, 2s-1$  as irreducible element of length
$s$.

Let $\phi_n$ be a canonical homomorphism of $GD(K)$ onto $GD(n,K)$.

The following proposition follows from the results on the girth of
  previous section. Now it is very important that $K=F_q$

\begin{proposition}
The order of each nonidentical element of $GD(F_q)$ is an infinity.
Let $g \in GD(F_q)$ be a regular element of length $l(g)=k$, then
the order of $g_n= \phi_n(g)$, where $k \le [n+5]/2 $, is bounded
below by $[n+5]/4k$ The sequence $g_n$ is a family of stable
elements.
\end{proposition}

So element $h=\tau^{-1}h^{-1}g_nh\tau$, where $\tau \in AGL_n(K)$,
$h\in DG(n,K)$ is an element for which $h^{-1}g_nh$ is a cubical
map, can be used as the base for Diffie-Hellman algorithm as above
for $K=F_q$.

\section{On the regular directed graph with special colouring}

Directed graph is an irreflexive binary relation $\phi \subset
V\times V$, where V is the set of vertices.

Let us introduce two sets
$$id(v)=\{x\in V|(a,x)\in \phi\},$$
$$od(v)=\{x\in V|(x,a)\in \phi\}$$
as sets of inputs and outputs of vertex v. Regularity means the
cardinality of these two sets (input or output degree) are the same
for each vertex.

Let $\Gamma$ be regular directed graph, $E(\Gamma)$ be the set of
arrows of graph $\Gamma$. Let us assume that additionally we have a
colouring function i.e. the map $\pi : E \rightarrow M$ onto set of
colours $M$  such that for each vertex $v \in V$ and $\alpha \in M$
there exist unique neighbour $u \in V$ with property
$\pi((v,u))=\alpha$ and the operator  $N_{\alpha}(v):=N(\alpha, v)$
of taking the neighbour $u$  of a vertex $v$ within the arrow $
v\rightarrow u$ of colour $\alpha$ i a bijection. In this case we
refer to $\Gamma$ as {\em rainbow-like graph}.

   For each string of colours $(\alpha_1, \alpha_2, \ldots, \alpha_m)$,
${\alpha_i}\in M$ we can generate a permutation $\pi$ which is a
composition $N_{\alpha_1}\times N_{\alpha_2}\times\cdots \times
N_{\alpha_m}$ of bijective maps $N_{\alpha_i}:V(\Gamma)\rightarrow
V(\Gamma)$. Let us assume that the map $u\rightarrow N_\alpha(u)$ is
a bijection. For given vertex $v\in V(\Gamma)$  the computation
$\pi$ corresponds to the  chain in the graph:
$$v\rightarrow v_1=N(\alpha_1, v)\rightarrow v_2=N(\alpha_2,
v_1)\rightarrow \cdots\rightarrow v_n=N(\alpha_m, v_{m-1})=v'.$$ Let
$G_{\Gamma}$ be the group generated by permutations $\pi$ as above.

 E.Moore \cite{Moore} used the term {\em tactical configuration} of order $(s, t)$
for biregular bipartite simple graphs with bidegrees $s+1$ and
$r+1$. It corresponds to the incidence structure with the point set
$P$, line set $L$ and symmetric incidence relation $I$. Its size can
be computed as $|P|(s+1)$ or $|L|(t+1)$.

Let $F=\{ (p, l)| p \in P , l \in L, pIl \}$ be the totality of
flags for the tactical configuration with partition sets $P$ (point
set)  and  $L$ (line set) and incidence relation $I$. We define the
following irreflexive binary relation $\phi$ on the set $F$: Let
$(P, L, I)$ be the incidence structure corresponding to regular
tactical configuration of order $t$.

Let $F_1 = \{(l, p)| l\in L, p \in P, lIp \}$ and $F_2= \{ [l, p] |
l \in L, p \in P, lIp \}$ be two copies of the totality of flags for
$(P, L, I)$. Brackets and parenthesis allow us to distinguish
elements from $F_1$ and $F_2$.  Let $DF(I)$ be the directed graph
(double directed flag graph)  on the disjoint union of $F_1$ with
$F_2$ defined by the following rules

$(l_1, p_1)\rightarrow [l_2, p_2]$ if and only if $p_1=p_2$ and $l_1
\ne l_2$,

$[l_2, p_2]\rightarrow (l_1, p_1)$ if and only if $l_1=l_2$ and $p_1
\ne p_2$.

\section{Construction of new stable
groups corresponding to rainbow like graphs}

Let us consider double directed graph $DD(n, K)$ for the bipartite
graph $D(n, K)$ and infinite double directed flag graph $DD(K)$ for
$D(K)$($DD(K) $) defined over the commutative ring $K$, Let
$N=N_{\alpha, \beta} (v)$ be the operator of taking the neighbor
alongside the output arrows of colours $\alpha, \beta \in {\rm
Reg(K)}$ of vertex $v \in F_1 \cup F_2$ by the following rule. If $v
= <(p), [l]> \in F_1$ then $N(v)=v'=[[l], (p')] \in F_2$, where  the
colour of $v'$ is $\alpha = p'_{1, 0}-p_{1,0}$, if $v= [[l], (p)]
\in F_2$ then $N(v)=v'=<(p), [l']> \in F_1$, where  the colour of
$v'$ is $\beta= l'_{1, 0}-l_{1,0}$.

Let us consider the elements $Z(\alpha, \beta)=N_{\alpha, 0}N_{0,
\beta}$. It moves $v \in F_1$ into $v'\in F_1$ at distance two from
$v$ and fixes each $u \in F_2$. Notice that $Z(\alpha,
\beta)Z(-\alpha, -\beta)$ is an identity map.

We consider the group $GF(n+1,K)$ ($GF(K)$, respectively) generated
by all transformations $Z(\alpha, \beta)$ for nonzero $\alpha, \beta
\in K$ acting on the variety $F_1=K^{n+1}$ ($K^{\infty}$).

\begin{theorem}
Sequence of subgroups $GF(n,K)$ of Cremona group $C(n,K)$ form a
family of subgroups of degree 3.

\end{theorem}
\textbf{Proof}

In the first step we connect a point with a line to get two sets of
vertices of new graph:
$$F=\{\langle (p),[l]\rangle ~ | ~ (p)I[l]\}~~~\cong K^{n+1} $$
$$F^{'}=\{\{[l],(p)\}~ | ~ [l]I(p)\}~~~\cong K^{n+1}.$$

Now we define the following relation between vertices of the new
graph:
$$\langle (p),[l]\rangle R \{[l^{'}],(p^{'})\} ~~\Leftrightarrow
~~[l]=[l^{'}] ~~\& ~~p_1-p_1^{'}\in  K $$
$$\{[l^{'}],(p^{'})\} R \langle (p),[l]\rangle ~~\Leftrightarrow
~~(p^{'})=(p) ~~\& ~~l_1^{'}-l_1 \in  K$$ Our key will be
$\alpha_1,\alpha_2,\ldots,\alpha_n$, such that $\alpha_i\in Reg K$.

  As a first vertex we take
$$\{[l],(p)\}=(l_1,l_{1,1},l_{1,2},\ldots,l_{i,j},p_1)$$ (our
variables) . Using the above relation we get get next vertex:
$$\langle (p)^{(1)},[l]^{(2)}\rangle =
(p_1,p_{1,1}^{(1)},\ldots,p_{i,j}^{(1)},l_1+\alpha_1)$$ with
coefficients of degree 2 or 3, where
$$
\begin{array}{l}
     p_{1,1}^{(1)}=l_{1,1}-l_1p_1, ~~~~deg=2\\
     p_{1,2}^{(1)}=l_{1,2}-l_{1,1}p_1~~~~deg=2\\
     p_{2,1}^{(1)}=l_{2,1}-l_1(l_{1,1}-l_1p_1)~~~~deg=3 \\
     p_{i,i}^{'(1)}=l_{i,i}^{'}-p_1l_{i,i-1}~~~~deg=2\\
     p_{i,i+1}^{(1)}=l_{i,i+1}-p_1l_{i,i}~~~~deg=2\\
     p_{i,i}^{(1)}=l_{i,i}-l_1(l_{i-1,i}-p_1l_{i-1,i-1})~~~~deg=3\\
     p_{i+1,i}^{(1)}=l_{i+1,i}-l_1(l_{i,i}^{'}-p_1l_{i,i-1})~~~~deg=3\\
\end{array}$$

Similarly we get the third vertex:

$$\{[l]^{(2)},(p)^{(3)}\}=(l_1+\alpha_1,l_{1,1},\ldots,l_{i,j},p_1+\alpha_2)$$
also with coefficients of degree 2 or 3, where
$$
\begin{array}{l}
     l_{1,1}^{(2)}=l_{1,1}+l_1p_1, ~~~~deg=2\\
     l_{1,2}^{(2)}=l_{1,2}+\alpha_1p_1^2~~~~deg=2\\
     l_{2,1}^{(2)}=l_{2,1}+\alpha_1p_{1,1}^{(1)}~~~~deg=2 \\
     l_{i,i}^{(2)}=l_{i,i}+\alpha_1p_{i-1,i}^{(1)}~~~~deg=2\\
     l_{i+1,i}^{(2)}=l_{i+1,i}+\alpha_1p_{i,i}^{'(1)}~~~~deg=2\\
     l_{i,i}^{'(2)}=l_{i,i}^{'}+\alpha_1p_1p_{i-1,i-1}^{'(1)}~~~~deg=3\\
     l_{i,i+1}^{(2)}=l_{i,i+1}+\alpha_1p_1p_{i-1,i}^{(1)}~~~~deg=3\\
\end{array}$$

Let us represent:
$$p_1^{(2k-1)}=p_1+\alpha_2+\alpha_4+\ldots+\alpha_{(2k-2)}=p_1^{(2k-3)}+\alpha_{(2k-2)}$$
$$l_1^{(2k)}=l_1+\alpha_1+\alpha_3+\ldots+\alpha_{(2k-1)}=l_1^{(2k-2)}+\alpha_{(2k-1)}$$

Assume that the following vertices:
$$\langle (p)^{(2k-1)},[l]^{(2k)}\rangle =
(p_1^{(2k-1)},p_{1,1}^{(2k-1)},\ldots,p_{i,j}^{(2k-1)},l_1^{(2k)})$$
$$\{[l]^{(2k)},(p)^{(2k+1)}\}=(l_1^{(2k)},l_{1,1}^{(2k)},\ldots,l_{i,j}^{(2k)},p_1^{(2k+1)})$$
have degrees:
\begin{displaymath}
\deg p_{i,j}^{(2k-1)}(l_1,l_2,\ldots,l_k,p_1)= \left\{
\begin{array}{ll}
2, & \quad {(i,j)}={(i,i)}^{'}~~ \textrm{or} ~~{(i,j)}={(i,i+1)},\\
3, & \quad {(i,j)}={(i,i)}~~ \textrm{or} ~~{(i,j)}={(i+1,i)} \quad
\end{array}
\right.
\end{displaymath}

\begin{displaymath}
\deg l_{i,j}^{(2k)}(l_1,l_2,\ldots,l_k,p_1)= \left\{
\begin{array}{ll}
3, & \quad {(i,j)}={(i,i)}^{'}~~ \textrm{or} ~~{(i,j)}={(i,i+1)},\\
2, & \quad {(i,j)}={(i,i)}~~ \textrm{or} ~~{(i,j)}={(i+1,i)} \quad
\end{array}
\right.
\end{displaymath}

Now we would like to find out degrees of polynomials of the vertices
$\langle (p)^{(2k+1)},[l]^{(2k+2)}\rangle$ and
$\{[l]^{(2k+2)},(p)^{(2k+3)}\}$.

We have the components of the vertices with corresponding degrees: :
$$
\begin{array}{l}

p_{i,i}^{'(2k+1)}=p_{i,i}^{'(2k-1)}-\alpha_{2k}l_{i,i-1}^{(2k)}~~~~deg=2\\

p_{i,i+1}^{(2k+1)}=p_{i,i+1}^{(2k-1)}-\alpha_{2k}l_{i,i}^{(2k)}~~~~deg=2\\

p_{i,i}^{(2k+1)}=p_{i,i}^{(2k-1)}+\alpha_{2k}l_1^{(2k)}l_{i-1,i-1})^{(2k)}~~~~deg=3\\

p_{i+1,i}^{(2k+1)}=p_{i+1,i}^{(2k-1)}+\alpha_{2k}l_1^{(2k)}l_{i,i-1}^{(2k)}~~~~deg=3\\
\end{array}$$
and
$$
\begin{array}{l}

l_{i,i}^{(2k+2)}=l_{i,i}^{(2k)}+\alpha_{2k+1}p_{i-1,i}^{(2k+1)}~~~~deg=2\\

l_{i+1,i}^{(2+2)}=l_{i+1,i}^{(2k)}+\alpha_{2k+1}p_{i,i}^{'(2k+1)}~~~~deg=2\\

l_{i,i}^{'(2+2)}=l_{i,i}^{'(2k)}+\alpha_{2k+1}p_1^{(2k+1)}p_{i-1,i-1}^{'(2k+1)}~~~~deg=3\\

l_{i,i+1}^{(2+2)}=l_{i,i+1}^{(2k)}+\alpha_{2k+1}p_1^{(2k+1)}p_{i-1,i}^{(2k+1)}~~~~deg=3\\
\end{array}$$

Hence using the induction we got:
\begin{displaymath}
\deg p_{i,j}^{(2k+1)}(l_1,l_2,\ldots,l_k,p_1)= \left\{
\begin{array}{ll}
2, & \quad {(i,j)}={(i,i)}^{'}~~ \textrm{or} ~~{(i,j)}={(i,i+1)},\\
3, & \quad {(i,j)}={(i,i)}~~ \textrm{or} ~~{(i,j)}={(i+1,i)} \quad
\end{array}
\right.
\end{displaymath}

\begin{displaymath}
\deg l_{i,j}^{(2k+2)}(l_1,l_2,\ldots,l_k,p_1)= \left\{
\begin{array}{ll}
3, & \quad {(i,j)}={(i,i)}^{'}~~ \textrm{or} ~~{(i,j)}={(i,i+1)},\\
2, & \quad {(i,j)}={(i,i)}~~ \textrm{or} ~~{(i,j)}={(i+1,i)} \quad
\end{array}
\right.
\end{displaymath}
  Finally using the affine transformation in the same way as in \cite{Wrob},
  independently from the length of the password we get the
  polynomials of degree 3.

Canonical graph homomorphisms $\omega_n: DD(n, K) \rightarrow
DD(n-1, K)$ can be naturally expanded to group homomorphism
$GF(n+1,K)$ onto $GF_{n}(K)$. It means that group $GF(K)$ is a
projective limit of $GF(n,K)$. Let $\delta_n$ be a canonical
homomorphism of $GF(K)$ onto $GF(n,K)$.

Let ${\rm Reg}(K)$ be the totality of regular elements of $K$ i. e.
non zero divisors.  We may consider the restriction
$\widetilde{DD(n, K)}$ of the graph $DD(n, K)$ via the following
additional condition.

$$\langle (p),[l]\rangle R \{[l^{'}],(p^{'})\} ~~\Leftrightarrow
~~[l]=[l^{'}] ~~\& ~~p_1-p_1^{'}\in {\rm Reg}(K) $$
$$\{[l^{'}],(p^{'})\} R \langle (p),[l]\rangle ~~\Leftrightarrow
~~(p^{'})=(p) ~~\& ~~l_1^{'}-l_1 \in {\rm Reg}(K)$$. We restrict
operators $N_{\alpha, \beta}$ and $Z(\alpha, \beta)$ simply by
adding the restrictions $\alpha, \beta \in {\rm Reg}(K)$. Let
$Q_n=Q(n, K)$ be the restricted group and $Q=Q(K)$ is a projective
limit of $Q(n, K)$, $n \rightarrow \infty$.

In \cite{Varna}, \cite{Max} was shown that the projective limit of
graphs $\widetilde{DD(n, K)}$ is acyclic graph and the length of
minimal directed cycle in $\widetilde{DD(n, K)}$ is bounded below by
$[n+5]/2$. It means that we get the following statement.

\begin{proposition}
The order of each nonidentical element of $Q(K)$ is infinity. Let $g
\in Q(K)$ be an element of length $l(g)=k$, then the order of its
projection $g_n= \delta_n(g)\in Q_n$, where $k \le [n+5]/2$, is
bounded below by $[n+5]/2k$ The sequence $g_n$ forms a family of
stable elements of increasing order. \end{proposition}

Theorem 1 follows immediately from theorem 4 and proposition 5.

\section{On the time evaluation for the public rule}

 Recall, that we combine a graph
 transformation $N_l$
 with two affine transformation $T_1$ and $T_2$.
 Alice can use $T_1N_lT_2$ for the construction of the following
public map of

$$y=(F_1(x_1, \dots , x_n), \dots , F_n(x_1, \dots , x_n))$$

$F_i(x_1, \dots , x_n)$ are polynomials of $n$ variables written as
the sums of monomials of kind $x_{i_1}^{m_1}x_{i_2}^{m_2}
x_{i_3}^{m_3}$ with the coefficients from $K=F_q$, where
$i_1,i_2,i_3\in{1,2,\dots, n}$ and $m_1, m_2, m_3$ are positive
integer such that $m_1+m_2+m_3\leq 3$. As we mentioned before the
polynomial equations $y_i=F_i(x_1,x_2,\dots,x_n),~i=1,2\dots n$,
which are made public, have the degree $3$. Hence the process of an
encryption and a decryption can be done in polynomial time $O(n^4)$
(in one $y_i,~ i=1,2\ldots,n$ there are $2(n^3-1)$ additions and
multiplications). But the cryptoanalyst Cezar, having only a formula
for $y$, has a very hard task to solve the system of $n$ equations
of $n$ variables of degree $3$. It is solvable in exponential time
$O(3^{n^4})$ by the general algorithm based on Gr{\"o}bner basis
method. Anyway studies of specific features of our polynomials could
lead to effective cryptanalysis. This is an open problem for
specialists.

We have written a program for generating a public key and for
encrypting text using the generated public key. The program is
written in C++ and compiled with the Borland bcc 5.5.1 compiler.

We use a matrix in which all diagonal elements equal 1, elements in
the first row are non-zero and all other elements are zero as $A$,
identity matrix as $B$ and null vectors as ${\rm c}$ and ${\rm d}$.
In such a case the cost of executing affine transformations is
linear.

The table \ref{tab:pkgen} presents the time (in milliseconds) of the
generation of the public key depending on the number of variables
($n$) and the password length ($p$).

\begin{table*}[tbp]
\caption{Time of public key generation} \label{tab:pkgen}
\begin{center}
\begin{tabular}{|c||r|r|r|r|r|r|}
\hline
& $p=10$ & $p=20$ & $p=30$ & $p=40$ & $p=50$ & $p=60$ \\
\hline \hline
$n=10$ & 15 & 15 & 16 & 32 & 31 & 32 \\
\hline

$n=20$ & 109 & 250 & 391 & 531 & 687 & 843 \\
\hline
$n=30$ & 609 & 1484 & 2468 & 3406 & 4469 & 5610 \\
\hline

$n=40$ & 2219 & 7391 & 12828 & 18219 & 24484 & 29625 \\
\hline
$n=50$ & 5500 & 17874 & 34078 & 49952 & 66749 & 82328 \\
\hline
$n=60$ & 12203 & 42625 & 87922 & 138906 & 192843 & 242734 \\
\hline
$n=70$ & 22734 & 81453 & 169250 & 286188 & 405500 & 536641 \\
\hline
$n=80$ & 46015 & 165875 & 350641 & 619921 & 911781 & 1202375 \\
\hline
$n=90$ & 92125 & 332641 & 708859 & 1262938 & 1894657 & 2525360 \\
\hline
$n=100$ & 159250 & 587282 & 1282610 & 2220610 & 3505532 & 4899657 \\

\hline
\end{tabular}
\end{center}
\end{table*}

The table \ref{tab:pkenc} presents the time (in milliseconds) of
encryption process depending on the number of bytes in plaintext
($n$) and the number of bytes in a character ($w$).

\begin{table}[tbp]
\caption{Time of encryption} \label{tab:pkenc}
\begin{center}
\begin{tabular}{|c||r|r|r|}
\hline & $Z_{2^8}$ & $Z_{2^{16}}$ & $Z_{2^{32}}$ \\
\hline \hline
$n=20$ & 16 & 0 & 0 \\
\hline
$n=40$ & 265 & 47 & 15 \\
\hline
$n=60$ & 1375 & 188 & 15 \\
\hline
$n=80$ & 3985 & 578 & 47 \\
\hline
$n=100$ & 10078 & 1360 & 125 \\
\hline
\end{tabular}
\end{center}
\end{table}

\end{document}